# XMM–Newton Data Processing for Faint Diffuse Emission

## Proton Flares, Exposure Maps and Report on EPIC MOS1 Bright CCDs Contamination

J. Pradas and J. Kerp

Radioastronomisches Institut der Universität Bonn, Auf dem Hügel 71, 53121 Bonn, Germany

**Abstract.**
We present a study of the *in–flight* performance of the XMM–Newton EPIC MOS and pn detectors, with focus on the influence of *proton flares* and *vignetting* on the data. The very wide range in the conditions of our sample of observations, in terms of exposure length and background intensities, allows the detection of a wide range in the spectra of the proton flares, in contrast to the hard–spectrum flares proposed by Lumb et al. (2002) or Read & Ponman (2003). We also find an up to now unreported contamination in the low energy regime ($E \leq 0.5\,\mathrm{keV}$) of the MOS1 observations, consisting of a significant increase in the measured intensities in two CCDs at the edges of the detector. This contamination yields in *bright CCDs* in the observations. Its effect must be taken into account for the study of sources detected in the affected CCDs. With respect to vignetting, we present *in–flight* exposure maps and we propose a method to repeat this calculation for user–definable energy bands. All the results presented here, have the goal to enable the study of very faint extended sources with XMM–Newton, like nearby galactic X–ray halos or the soft X–ray background.

## 1. Introduction

The investigation of low surface brightness sources in X–ray astronomy requires very precise data reduction methods. In the case of the XMM–Newton observatory, several authors (e.g., Lumb et al. 2002; Read & Ponman 2003; Katayama et al. 2004) have addressed the data reduction problem in the standard data analysis procedure applied to the case of faint and extended sources. Most of the effort to improve the standard data reduction tools has been focused on the development of temporal filters for the mitigation of the so–called *proton flares* and minimizing the instrumental noise with high accuracy (see, e.g. Lumb et al. 2002). Read & Ponman (2003) published *background maps* to correct the observations for *vignetting*, which is of special importance for the analysis of extended sources. But the contribution of the cosmic X–ray background, originating from the Local Hot Bubble, the Galactic Halo and the extragalactic X–ray background, has not been systematically treated for XMM–Newton as it has been in case of the ROSAT mission (e.g., Kerp et al. 1999; Snowden et al. 1998; Pradas et al. 2003). Because XMM–Newton is, in principle, able to detect very faint signals, all systematic effects must be well understood and reliable methods to eliminate their contributions have to be developed. All this effort should lead to an *absolute calibration* of the X–ray data which allows the study in detail of very faint extended sources (e.g., warm/hot intergalactic medium (WHIM), X–ray halos of nearby galaxies).

Here we present our newly developed data reduction method for the analysis of diffuse X–rays observations with the XMM–Newton EPIC MOS and pn detectors. In order to develop this method, we made extensive use of the XMM–Newton Science Archive[1] and carried out the *in–flight* background analysis for XMM–Newton with largest accumulated exposure time at present (about 3.8 Ms). In this work, we focus on the *FullFrame* mode, although the developed method can be straightforwardly applied to the remaining modes.

We also report on an overestimation in the measured X–ray intensity in the softest energy regime ($0.2\,\mathrm{keV} \leq E \leq 0.5\,\mathrm{keV}$) for CCDs 2 and 5 of the EPIC–MOS 1 camera. These two CCDs occasionally show up with up to a factor of two higher X–ray background intensity than the other CCDs. In Sect. 5 we will see that about 15% of the observations in our database are affected by this effect in CCD 5. In case of CCD 2, more than 50% of the observations present contamination detectable with a first visual inspection of the data.

The structure of this paper is as follows: In Sect. 2 we present the X-ray data and its calibration. In Sect. 3, we present

---

*Send offprint requests to*: J. Pradas,
e-mail: `jpradas@astro.uni-bonn.de`

[1] *http://xmm.vilspa.esa.es/external/xmm_data_acc/xsa/index.shtml*

our new data reduction method with emphasis on the detection of proton flares (see Sect. 3.2). In Sect. 4, we explain the construction of the new *background maps* based on the in–flight performance of XMM–Newton. In Sect. 5, we report on the detection of *bright CCDs* in several observations of the EPIC–MOS 1 camera. In Sect. 6, we present our conclusions.

## 2. XMM-Newton Raw and Calibrated Data

The database that we have used for this investigation is separated in two types of observations. On one hand, we make use of as many public XMM–Newton Science Archive observations as possible (see Tabs. 3 to 5) in order to have a significant sample of the general response of the instruments in a large variety of configurations (e.g,. high and low background levels or absorber column densities). On the other hand, we make use of a set of own observations (see Tab. 6) towards selected regions in the sky, for which there is a special importance of the contribution of the XRB. For example, a high contrast region towards the Draco Nebula area is included in observation 0110660801.

### 2.1. Observations from the XMM–Newton Science Archive

Using the XMM–Newton Science Archive, we searched for all available observations in which the EPIC cameras worked in the *FullFrame* mode. This choice is based on our interest to develop tools for investigating the diffuse soft X–ray emission ($E \leq 2\,\mathrm{keV}$) which requires the use of the maximal detector surface.

In order to maximize the background signal of the sample, observations with very bright point sources were excluded from our data selection. We also excluded observations with less than 8 ks exposure time. This criterion assures an expected background rate of at least $1\,\mathrm{cts\,arcmin^{-2}}$ for the soft energy regime, defining a set with acceptable statistical significance. The derived value of 8 ks is based on the ROSAT all–sky survey (RASS) data, taken as a first estimation of the level of the cosmic X–ray background (XRB) towards the fields of interest. In fact, the minimal X–ray intensity for the XRB among the observations in Tabs. 3 to 6 of $119 \cdot 10^{-6}\,\mathrm{cts\,s^{-1}\,arcmin^{-2}}$ observed by ROSAT in the R2 energy band corresponds to $\simeq 120 \cdot 10^{-6}\,\mathrm{cts\,s^{-1}\,arcmin^{-2}}$ in one EPIC MOS detector ($E \leq 2\,\mathrm{keV}$) with the *medium* filter ($1\,\mathrm{cts\,arcmin^{-2}}$ expected in about 8 ks). Note, that the differences in effective area between both mirrors must be considered for this calculation.

### 2.2. Selected Fields for the Investigation of the XRB

Tab. 6 shows a selection of fields of interest where, in some cases, the RASS revealed large intensity contrasts in the SXRB (X–ray shadows of interstellar clouds, like in Burrows & Mendenhall (1991), for observations 0110660801 and 0110662601 of Tab. 6). In the remaining fields of Tab. 6, no attenuation of the Galactic X–ray Halo emission by the H I "clouds" in the fields was detected by the analysis of the RASS (e.g. HVC complexes in Pradas et al. (A&A submitted) for observation 0110660401 of Tab. 6). These fields are also included in the sample in order to test the validity of the results of our method. All fields were observed with an accumulated integration time of at least 10 ks to assure good photon statistics (see above). These fields are also included in our study of the *proton flare filtering* and *vignetting* of in XMM–Newton (see following Sects.).

With the two subsets presented in this Sect., we accumulated the largest database to the present (about 3.8 Ms) for the analysis of the XMM–Newton data calibration based on the *in–flight* performance of the satellite (see Sects. 3.2 and 4). The observations were calibrated with respect to the calibration database synchronized on October 24, 2002. For that purpose, we use the standard SAS 5.3.3 tasks grouped in the *pipeline* procedure called *emchain*. Further processing of the so produced *calibrated event lists*, was performed using a combination of SAS 5.3.3 and self–developed software (see Sect. 3).

## 3. Data Reduction

For the investigation of diffuse X-ray emission, it is necessary to have an *absolute calibration* of the observations. Therefore, we need a precise knowledge of all contaminating effects in the data. This is even more important in the case of the soft energy regime, since the signal is very faint in comparison to the sources of contamination.

In contrast to the RASS, there is no standard tool to perform such an absolute calibration for faint extended emission for XMM–Newton. There are some problems in the data reduction procedure that require the development of special tools to be solved. For example, Read & Ponman (2003) and Lumb et al. (2002) agree in the necessity for a reliable *proton flare filtering*, alternative to the method presented in the *ABC Guide to XMM–Newton Data Analysis* (Snowden et al. 2004) available at the web pages of the *High Energy Astrophysics Science Archive Research Center*[2], and propose slightly different methods in order to achieve this goal. Read & Ponman (2003) and Kappes et al., in prep, also deepen in the analysis of the XMM–Newton vignetting and propose methods to overcome the insufficiency of the standard data reduction tools to deal with this effect. In practice, the standard tools produce an artificial enhancement of the intensity at the edges of the detector that can be described as a *tunnel effect* (see Sect. 4 and Fig. 3).

Now, we present the method that we developed for the data reduction based on the *in–flight* performance of XMM-Newton, and which has the goal of allowing a reliable study of very faint diffuse emission. We emphasize in the effect of proton flares (Sect. 3.2) and vignetting (Sect. 4) and make use of the largest database for this kind of investigation today.

### 3.1. Selection of the Energy Bands

The first step that we performed is the splitting of the *event lists* in the energy bands compiled in Tab. 1. We focus on the study of soft X-rays. Consequently, we select *three* energy bands in the range $E \leq 2\,\mathrm{keV}$, from B1 to B3. We neglect the energy range $E \leq 0.2$ keV because of the uncertainties in the calibration of the data in this energy regime as explained in the *ABC Guide to XMM–Newton Data Analysis* and, e.g., by Read & Ponman (2003). Band B4 covers the energy range dominated by the extragalactic X-ray background (Hasinger et al. 2001). This background is very homogeneous in the B4 energy regime and its homogeneity can be used to test the validity of the data reduction methods applied. If the data reduction methods do not lead to quite homogeneous maps of the extragalactic background (for relatively low integration times), the validity of these correction methods is questioned. In practice, the extragalactic background can be used as a "reference" to normalize the intensities of the XRB (see Sect. 4), as has been done in previous investigations using ROSAT data (e.g., Kerp et al. 1999; Snowden et al. 1998; Pradas et al. 2003).

---

[2] *http://heasarc.gsfc.nasa.gov/docs/xmm/abc*

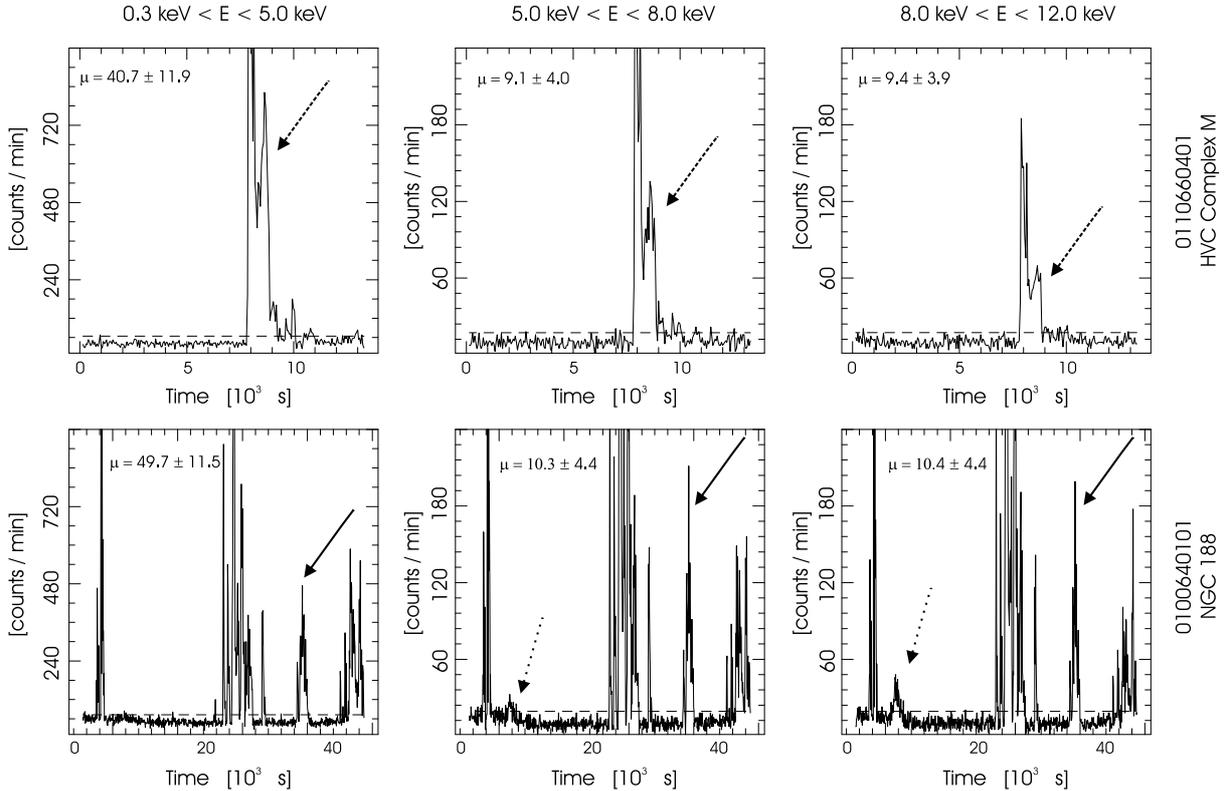

**Fig. 1.** Light curves of the EPIC MOS1 camera in the three C1, C2 and C3 energy bands (indicated on top of the panels) for two observations representative of the sample and indicated on the right side of the panels. The dashed lines in each panel marks the two sigma level used for the rejection of contaminated time intervals at the final iteration step (see Sect. 3.2). Mean count rates as determined by our programs are shown in the upper parts of the panels as $\mu$. The arrows (solid, dashed and dotted) point to the effect of three typical proton flares with significant differences in their respective spectra (see Sect. 3.2).

**Table 1.** Selected XMM-Newton energy bands.

| Band | $E_{min}$[KeV] | $E_{max}$[KeV] |
|---|---|---|
| B1 | 0.2 | 0.5 |
| B2 | 0.5 | 1.0 |
| B3 | 1.0 | 2.0 |
| B4 | 2.0 | 5.0 |
| C1 | 0.2 | 5.0 |
| C2 | 5.0 | 8.0 |
| C3 | 8.0 | 12.0 |

Energy bands C1, C2 and C3 are only related to source detection and *proton flare filtering*. The choice of these bands is justified in the following Sect. 3.2.

### 3.2. Proton Flares

The effect produced by *proton flares* (see, e.g. Read & Ponman 2003) in the observations performed by XMM–Newton is clearly visible in the light curves presented in Fig. 1. We now present a method to filter these events which is a fundamental step for a further analysis of diffuse emission with XMM–Newton data. In Marty et al. (2003) a compilation of alternative methods can be found. However, the methods compiled by Marty et al. cannot, in general, be applied in a fully automatic way, in contrast to our method presented here.

When the imaging instruments of the satellite cross interplanetary clouds of electrically charged particles, the count rate increases by up to several orders of magnitude. The low energy protons of these charged clouds are ejected from the Sun (Marty et al. 2003) and show a broad variety of X–ray spectra (see Fig. 1), with particle energies covering the entire energy coverage of the EPIC cameras. It has been proposed (Lumb et al. 2002) that proton flares are composed of several components with different spectra and turn on times. However, Lumb et al. (2002); Read & Ponman (2003) suggest that the most important fraction of the proton flares show a hard X–ray spectrum with their dominant contribution at energies higher than $E \geq 10 \, \text{keV}$ and, therefore, they focus their methods for proton flare filtering only on the hardest energy bands, corresponding approximately to the C3 band in Tab. 1 . We now revise this suggestion by systematically extending the search for proton flares to all energy bands noted in Tab. 1. The very broad bands C1, C2 and C3 are used because their high signal–to–noise ratio gives very stable count rates in the phases when the detectors are not being affected by proton flares. Finally, all three bands together cover the entire EPIC energy range without any overlap.

Today, there is no method available to predict the occurrence of proton flares. Therefore, their effect can only be corrected in a *post–observation* data analysis. An accurate method of detecting the presence, beginning and end of a proton flare is required to keep the longest usable observation time. We developed such a temporal–filter method based on an iterative algorithm with an user definable $\sigma$–level. This $\sigma$–level gives the minimal relative contribution of a proton flare to the total count rate necessary to flag out the corresponding time interval.

For each energy band, we compute the mean $\mu_i$ and standard deviation $\sigma_i$ of the count rate and search for observing intervals with a rate exceeding a threshold defined by the user, typically $\mu_i + 2\sigma_i$ or $\mu_i + 2.5\sigma_i$. These *bad time intervals* are flagged and $\mu_{i+1}$ and $\sigma_{i+1}$ are calculated for the remaining observing time. This iteration continues until the difference of the mean values of two consecutive iteration steps stays below the statistical uncertainty of the data ($\mu_i - \mu_{i+1} \leq \sqrt{\mu_{i+1}}$). This "stop condition" is generally fulfilled after less than five iterations. Then, we compute the intersection of the *good time intervals* obtained for the different energy bands and obtain the maximal observing time with *all* bands free of proton flare contamination. Then, good time intervals with a shorter duration than four minutes are rejected. **Bad time intervals shorter than two minutes are not rejected. These two constrains are obtained by trial and error from a large number of observations, although they can be changed in the software in order to "fine tune" it to specific situations. Based on our experience, we find that, in most cases, time intervals identified by the programs as good but with a shorter duration than four minutes, correspond to a decrease in the intensity of a flare between two epochs of higher intensity. However, the intensity decreased epoch is still contaminated by a flare. Time intervals identified as bad and shorter than two minutes, usually correspond to anomalous data in only a single energy band, and can be considered as good time intervals in all other energy bands. However, the here proposed interval duration constrains might be inadequate in some special cases and the corresponding parameters in the software should be changed accordingly. For an investigation of a large number of observations, like presented in this work, we can safely neglect the imprint on the final result of time intervals in the minute range.**

After the application of the proton flare detection on the complete event lists, a first source detection (see Sect. 3.4 for more details) is carried out. The obtained *source lists* are then used to filter the complete event lists and produce event lists containing only background events. With these new *source filtered* lists used as input, we invoke the proton flare filter a second time. By eliminating the contribution of point sources from the input for this second application of the filter, we mitigate the confusion created by sources at the step of calculating the background count rates for each observation. With that, we achieve an important improvement in the sensitivity of the proton flare detection , **with a difference of about 10% in the final total effective time of the sample. This difference is mainly because of flares, whose intensities are dominated in the total light curves of their observations by an important contribution of point sources. Then, after selecting the events corresponding to point sources, light curves of "pure" background events can be calculated. By using these light curves, the flares become statistically significant and can be detected by the programs.**

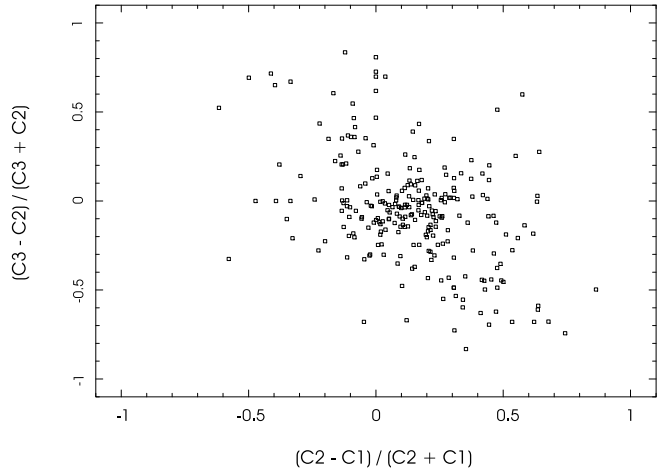

**Fig. 2.** Hardness ratios of the proton flares detected with our procedure for observations performed with the *medium* filter (see Sect. 3.2.1). With the energy bands C1, C2 and C3 we make use of the whole EPIC energy coverage. For clarity, we omit the error bars which vary in a broad range. This is because there are significant differences in the respective intensities of the proton flares represented in the Fig., from very faint to very bright which correspond to relatively large (about 35%) and small (about 10%) error bars respectively. We can see that the hardness ratios of the proton flares are wide spread in all ranges. In contrast to the suggestions by Read & Ponman (2003); Lumb et al. (2002), there is an important fraction of proton flares with *soft* and *medium* spectra in the lower half of the diagram.

In the final step, the *filtered event lists* corresponding to the good time intervals are calculated. From the initial 3.8 Ms exposure time included in the total sample, about 3.0 Ms effective time is left after the application of our filtering of proton flares for all EPIC cameras (see columns 5 to 9 in Tabs. 3 to 6), although the obtained effective exposure time for MOS2 is in general slightly larger than for MOS1.

### 3.2.1. Spectra of the Proton Flares

The arrows in Fig. 1 (solid, dashed and dotted) indicate typical examples of proton flares. We have chosen the same scale in the vertical axes of the diagrams for the two observations shown in order to allow a direct comparison of the hardness of the three marked flares. The dotted arrow corresponds to the flare with the hardest spectrum, with no significant contribution in the lowest energy regime (lower panels in Fig. 1). With a slightly softer spectrum than the previous flare, the solid arrow marks an example of a flare detected in all three energy regimes used. Finally, the dashed arrow indicates a proton flare with a very high contribution to C1 and a significantly softer spectrum than the other two flares of this example.

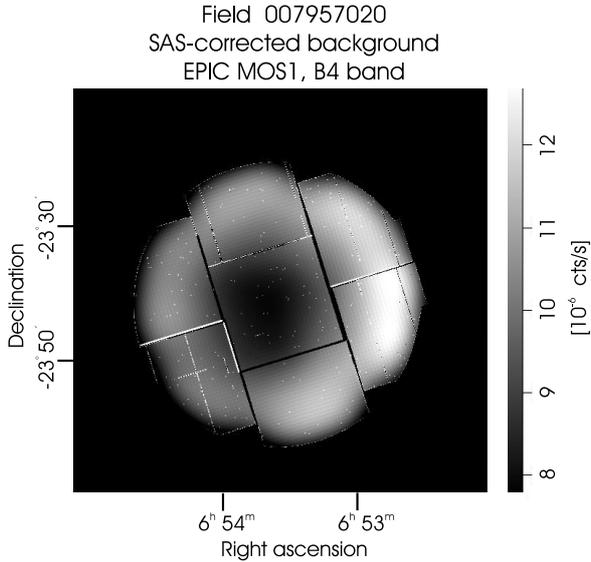

**Fig. 3.** X–ray background map towards the field indicated in the upper part of the panel, as obtained by the standard SAS data reduction tasks: The SAS–calibrated photon image is "source filtered" (see Sect. 3.4), smoothed and finally corrected by the exposure map calculated by the corresponding SAS task (*eexpmap*). The increase in intensity towards the rims of the detector (see *inverted tunnel effect* in Sect. 4) is created by the use of the task *eexpmap*. This resulting background map is unacceptable since, at the B4 band, the X–ray background is dominated by the extragalactic background (Read & Ponman 2003) which must yield in very homogeneous intensity distributions.

Equipped with our tools to automatically search for proton flares, we investigated a large number of XMM-Newton observations (see Tabs. 3 to 6). We found that proton flares show up in all XMM–Newton energy regimes with a broad variety of X–ray spectra. The results presented in Fig. 2 indicate that it is not safe to restrict the search for proton flares to the high energy regime. This is in opposition to the proposals by Lumb et al. (2002) or Read & Ponman (2003), which suggest a search for proton flares restricted to the $E \geq 10$ keV. Our results show that the search for proton flares should be extended to all energy regimes as part of the standard reduction of XMM–Newton data. **This is in agreement with the results recently published by Nevalainen et al. (2004), who also find soft energy flares in their investigation of the X–ray emission of galaxy clusters.**

### 3.3. Other Sources of Contamination

Following Read & Ponman (2003), there are other sources of systematic contamination in the XMM–Newton data that should be taken into account. Among these, *electronic noise* affects only the low energy regime $E \leq 0.3$ keV and is almost completely excluded from the analysis with our selection of energy bands (see Tab. 1). However, the energy range $0.2\,\text{keV} \leq E \leq 0.3\,\text{keV}$ was not excluded from our energy band selection in order to improve the statistics of the B1 band by increasing its bandwidth. This is justified because the most important part of the *electronic noise* in confined to the $E \leq 0.2$ keV regime *(ABC Guide to XMM–Newton Data Analysis)*. In the $0.2\,\text{keV} \leq E \leq 0.3\,\text{keV}$ regime, the electronic noise contamination can be neglected in comparison with the contribution of the XRB to a broad energy band like B1.

**The contribution of the un–rejected *cosmic–ray induced particle background* (CRB) is negligible (Lumb et al. 2002). The internal background is dominated by the contribution of the XRB in the energy range $E \leq 5$ keV (Fig. 8 from Lumb et al. (2002)).**

In the case of *fluorescent lines* in the energy regime from B1 to B4, like the Al–K line at $E \simeq 1.5$ keV (Lumb et al. 2002), we must consider that the contribution of narrow lines vanishes when we focus on the study of broad energy bands. Consequently, we exclude a treatment of this effect in our data reduction process. However, this assumption will be revised in the future, when the available database would permit a precise spectral investigation of the vignetting by making use of narrower energy channels as those used here (see Tab. 1) **Then, a higher statistical certainty in the data, could permit testing the spatial distribution of the fluorescent lines and a more specific analysis of the most intense lines.**

On the contrary, the contribution of the XRB is maximized for $E \leq 5.0$ keV and, in principle, it cannot be neglected in our study. In order to avoid a specific analysis of the XRB for each observation, which would dramatically increase the computation time for an automatic procedure applied to a large database like in the work at hand, we have chosen our sample to include a very wide range in Galactic coordinates (columns 2 and 3 in Tabs. 3 to 6). We also have a wide range in the column density of the photoelectric absorbing material ($0.6 \cdot 10^{20}\,\text{cm}^{-2} < N_{\text{H\,I}} < 909.1 \cdot 10^{20}\,\text{cm}^{-2}$). These selections ensure a broad range in the XRB intensities of our sample, as already confirmed by the RASS. Additionally, the large number of selected observations yields a rich variety in structures of absorber column density present in the individual fields. Then, by averaging all fields of the sample, the differences in the structure of the XRB for each field are compensated. **This will be of special importance in our study of the in–flight vignetting of XMM–Newton in Sect. 4 and the following. There, we will be interested in the radial slope of the vignetting function and the differences in intensities among the fields will be canceled by normalizing the images to the value at the center of the field.** Consequently, a specific study of the XRB in each field will be neglected in our investigation of the vignetting in XMM–Newton.

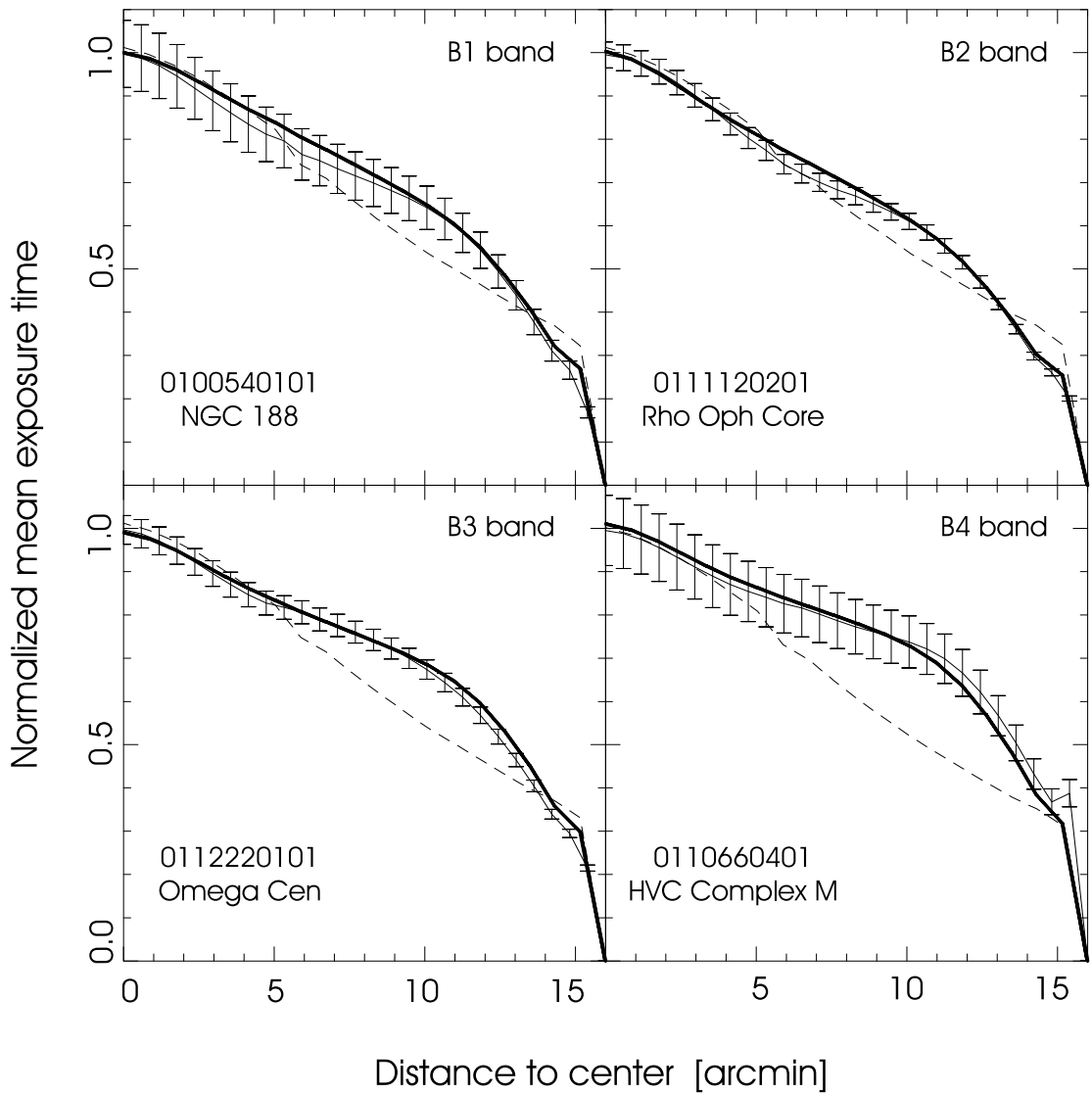

**Fig. 4.** Normalized mean exposure time vs. radial distance to the optical center of the EPIC MOS2 camera for the four energy bands shown at the upper right part of each diagram. The thick solid line shows our new developed exposure maps. To illustrate the quality of these exposure maps, we also show the background intensity distribution, without correction for vignetting, of four different observations (thin solid lines). The error bars are calculated based on the photon statistics of the inner 5 arcmin of the observations. The solid lines can be compared to the SAS calculated exposure maps (dashed lines) for the corresponding observations and bands. **The use of version 5.4.1 of the SAS software and a more modern calibration database (synchronization of March, 2004), as mentioned in Sect. 5, yields only slight variations in the shape of the obtained exposure maps.**

### 3.4. Source Detection

The analysis of the diffuse XRB also requires the rejection of the contribution of point sources to the data. Here, we invoked the SAS tasks *eboxdetect* (boxsize=5, likemin=10, nrun=4) and *emldetect* (mlmin=10, scut=0.9, ecut=0.68) for all observations in Tabs. 3 to 6 and for the energy band C1. This was performed in the *double run* mode proposed in the *ABC Guide to XMM–Newton Data Analysis* in which *eboxdetect* is run twice to obtain an appropriate background map which is then used as input for *emldetect*.

After completing the source detection, we make use of a self–developed tool to automatically eliminate the contribution of all point sources. This was done by the creation of so–called *cheese images* and posterior refilling of the gaps with the background level of a nearby *source–free* region. The refilling is necessary in order to avoid an artificial decrease in the intensity of our accumulated database towards the center of the detectors due to the higher source detection rate in the central area of the cameras. We explain this in some more detail in the following.

In Sect. 4 we will average the background maps obtained here and, for this aim, maps with a high number of sources close to the center of the field are unacceptable since they yield in a significant loss of XRB intensity in the inner 5 arcmin. A *refilling* step for every field is necessary. In this step, we fill each *hole* in the cheese images with an approximate background intensity value according to a source free region in its vicinity. For this purpose, we implemented an iterative procedure based on a Gaussian smoothing task (*asmooth* of SAS with smoothtype='simple') with $\sigma = 1.5$ arcmin. In each iteration, background areas (without sources) are kept constant. Then, the value obtained by the Gaussian filter for the areas with sources is used as input for the same areas in the next iteration step. In general at most five iteration steps are necessary until the difference between the areas with sources of two successive steps stays below the accuracy level of the data (see also Sect. 3.2).

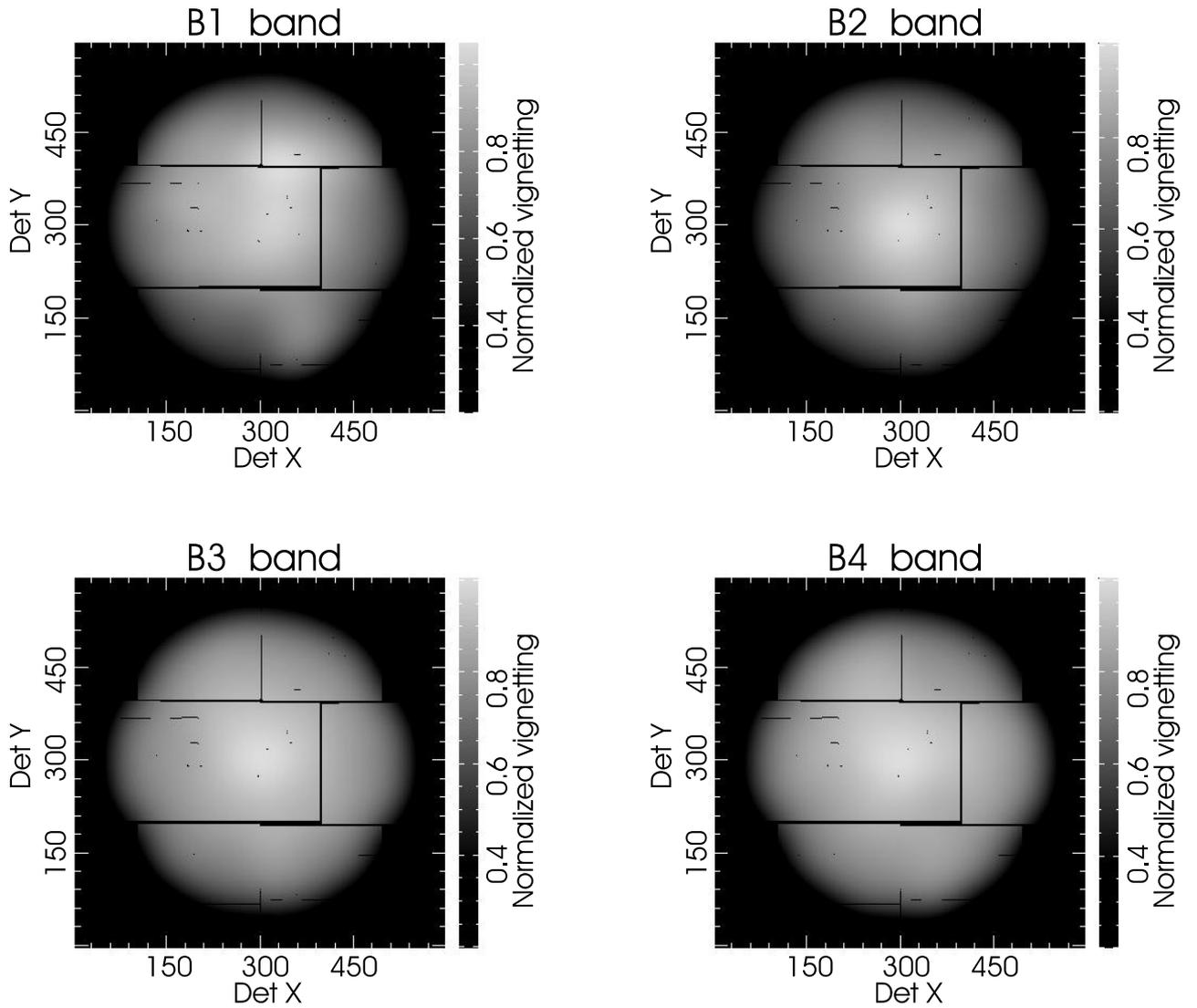

**Fig. 5.** Normalized exposure maps for the four energy bands indicated on the top of the panels obtained with our method applied to the EPIC MOS1 detector with the *medium* filter. The grids of each map are presented in detector coordinates. The gaps in the maps correspond to the detector mask. Apart from a small residual contamination from CCD2 (see Sect. 5) in the lower right part of the B1 band map, a good radial symmetry is present in all panels. The total effective exposure times used to calculate the maps of this Fig. are $t_{\rm eff} \simeq 0.45$ Ms for B1, $t_{\rm eff} \simeq 0.82$ Ms for B2 and $t_{\rm eff} \simeq 1.15$ Ms for bands B3 and B4. The differences in the total times are related to bright CCD contamination (see Sect. 5 and Fig. 9).

Our choice of a smoothing size of $\sigma = 1.5$ arcmin is justified with similar arguments as for the exposure times in Sect. 2. At least one X–ray photon is expected in each point of the new resolution grid with 1.5 arcmin angular spacing.

## 4. Exposure Maps

The final step in the data reduction process for diffuse emission involves the correction due to the *vignetting* of the XMM–Newton mirrors. Vignetting is an effect that depends on photon energy and causes a gradual decrease of sensitivity towards the rim of the FOV. This dependency stems from the reduction of effective area of the detectors with off–axis angle and photon energy. The standard method to correct for vignetting consists in dividing the observed maps through the so–called *exposure maps*.

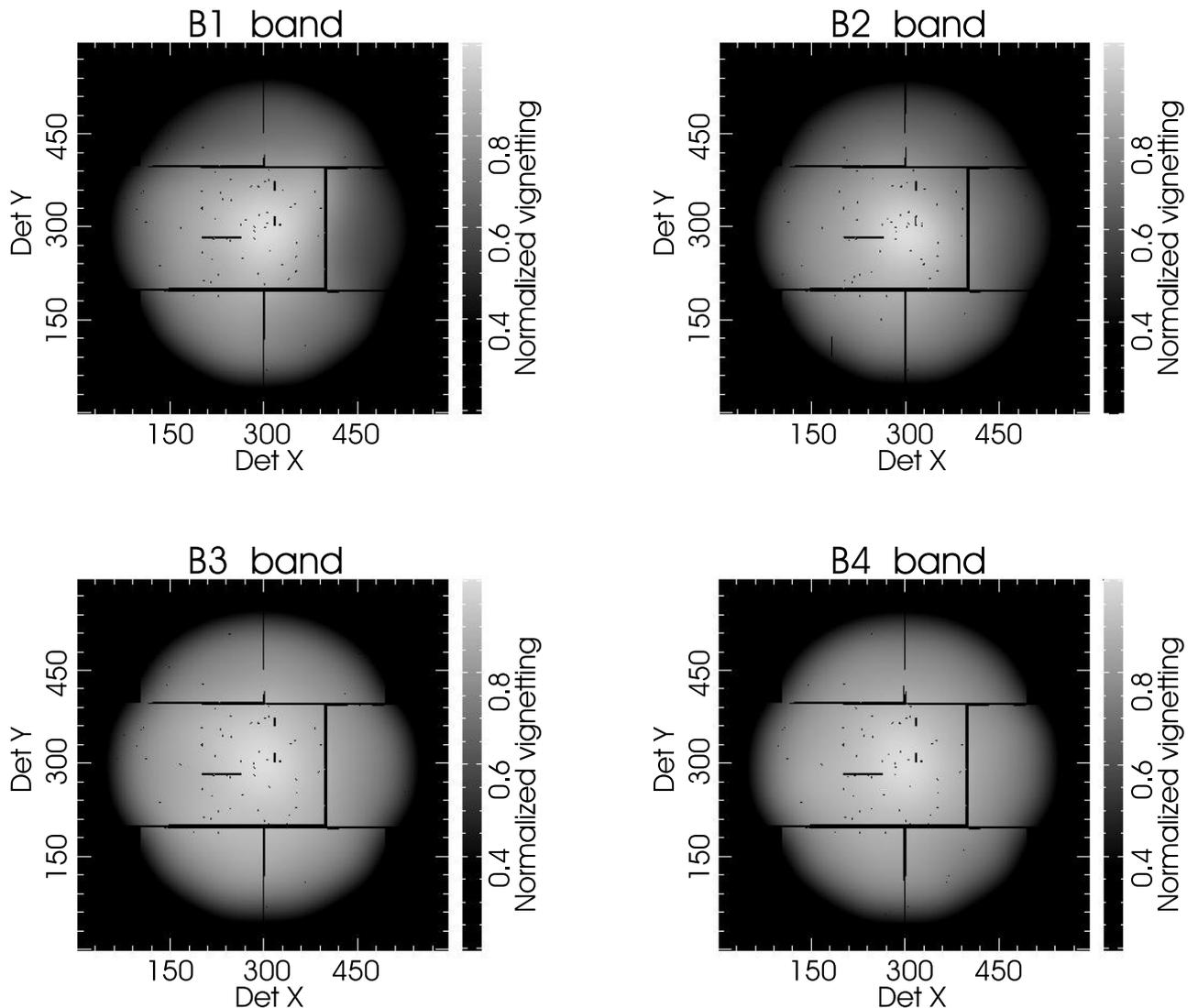

**Fig. 6.** Normalized exposure maps for the four energy bands indicated on the top of the panels obtained with our method applied to the EPIC MOS2 detector with the *medium* filter. The grids of each map are presented in detector coordinates. The gaps in the maps correspond to the detector mask. A good radial symmetry is present in all panels. The total effective exposure times used to calculate the maps of this Fig. are $t_{\rm eff} \simeq 1.15$ Ms in all cases.

We tested the SAS 5.3.3 tool *eexpmap* for the calculation of the exposure maps. We detected an overestimation of the correction towards the rims of the FOV (see Fig. 3), which yields final intensity maps with very bright rims (*inverted tunnel effect*). Read & Ponman (2003) faced this problem and proposed a different method to correct for vignetting based on the *in–flight* performance of XMM–Newton. We extended their investigation to a longer accumulated exposure time (from $\simeq 1.5$ Ms to $\simeq 3.8$ Ms) and take advantage of the improved data reduction tools presented in the preceeding sections.

We use the background maps explained in Sect. 3.4 as a weighted sample to calculate the mean background intensity distribution affected by vignetting as observed by XMM–Newton *in–flight*. The weighting used for observations in the sample is proportional to the effective observation time (after proton filtering) as shown in columns 7 to 9 in Tabs. 3 to 6. In all cases, the background maps are converted to *detector coordinates* (*XMM–Newton Users' Handbook*, Ehle et al. (2004), available at the VILSPA web pages[3])). Thus, and since the observations are summed up always in the same orientation (see Sect. 5), we can study not only the radial distribution of the mean exposure maps, which is valid only if there is radial symmetry in the vignetting, but the complete 2–D distribution.

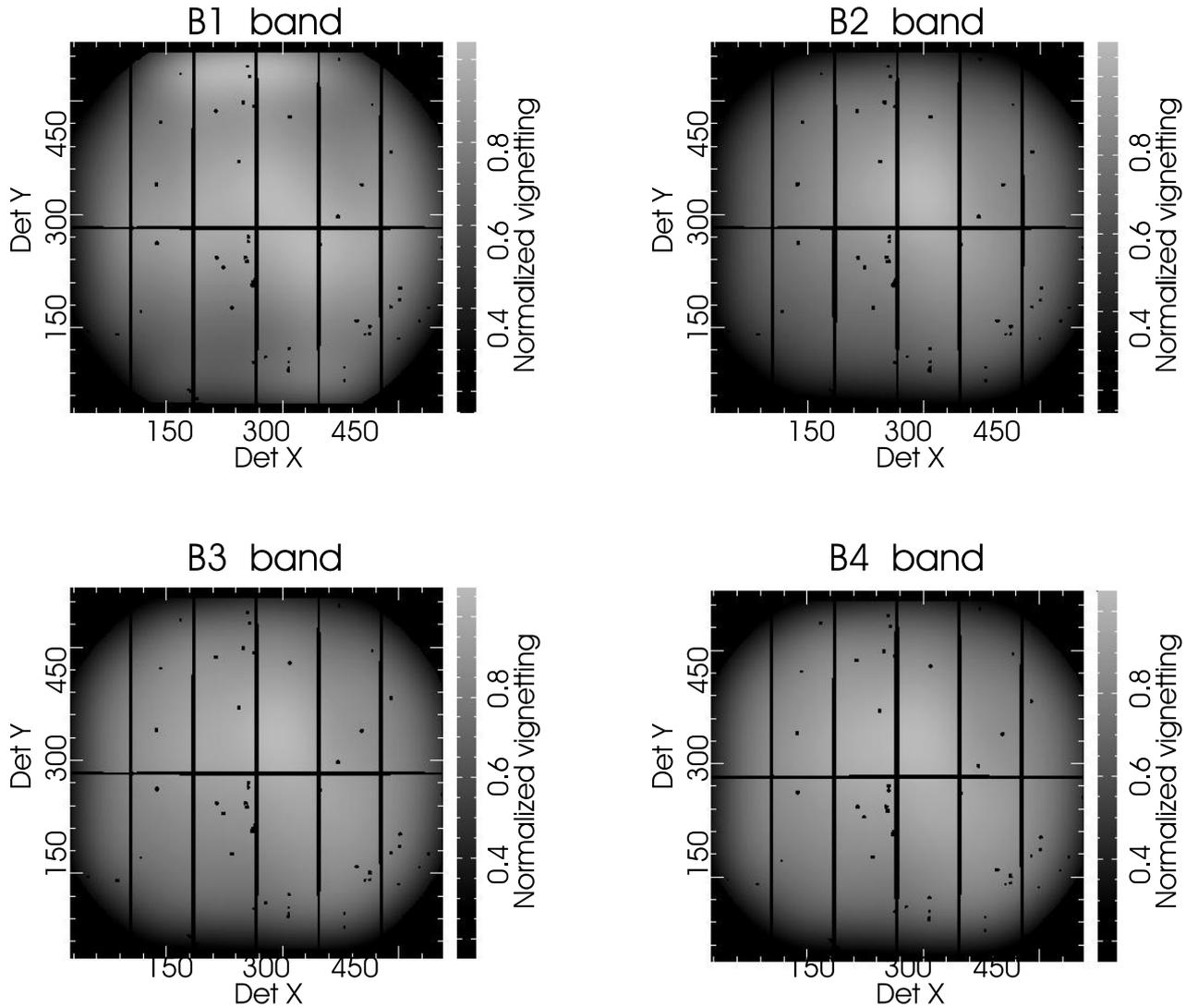

**Fig. 7.** Normalized exposure maps for the four energy bands indicated on the top of the panels obtained with our method applied to the EPIC pn detector with the *medium* filter. The grids of each map are presented in detector coordinates. The gaps in the maps correspond to the detector mask. Like in the case of MOS1 (see Fig. 5), there are clear unexpected asymmetries in the radial distribution of the B1 band vignetting. The total effective exposure times used to calculate the maps of this Fig. are $t_{\rm eff} \simeq 1.15$ Ms in all cases.

[3] *http://xmm.vilspa.esa.es/external/xmm_user_support/documentation*

In Fig. 4, we present our results for the medium and low energy regimes (B1 to B4, i.e. $0.2\,{\rm keV} < E < 5\,{\rm keV}$). We focus on these regimes because:

- B1 and B2 are of most importance for the investigation of the diffuse X–ray plasma like, e.g., galactic X–ray Halos or the warm/hot intergalactic medium. This is because the bulk of the X–ray emission of such plasmas, with temperatures in the million K regime, is radiated in the keV regime covered by bands B1 and B2.
- B3 and B4 are dominated by the extragalactic X–ray background and, since this emission is well–known and very

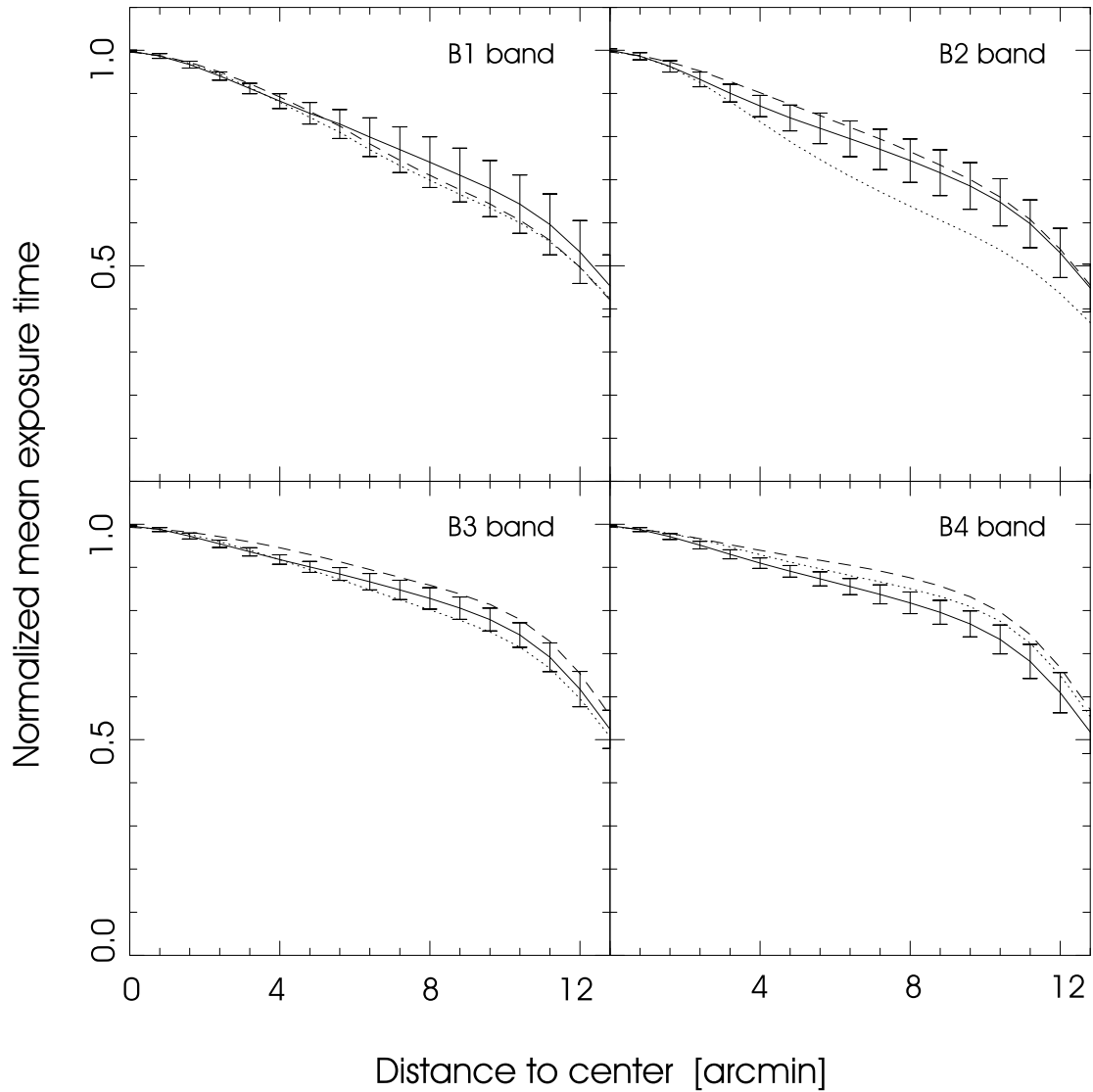

**Fig. 8.** Radial profiles of the exposure maps for the energy bands indicated in the upper right part of the diagrams in the cases of the *thin* (dotted line), *medium* (solid line) and *thick* (dashed line) filters. For clarity, only the error bars for the *medium* filter are shown. These are calculated based on the variation of the exposure across the rings used for the calculation of the radial profile. There are clear differences in the profiles for the different filters, like for the *thick* filter in the B2 panel.

homogeneous, these two bands are a good choice to perform the normalization of the intensity distribution across the fields of interest. **This is relevant for the use in individual observations of the exposure maps presented here. In that case, not only a normalized exposure map is needed but also a measure of the effective observing time is required to calculate count rates for the images. There, we can constrain the result to be compatible in the harder energy regimes with the values available from earlier missions (e.g. ROSAT). The normalization factor obtained for the harder energy regimes can be used to fix the corresponding values in lower energy regimes, which are more likely to be affected by additional sources of contamination.**

The presented method leads to a better agreement between the observations and our results than between observations and the respective exposure maps as calculated by *eexpmap*. Some examples, included in our database, are shown in Fig. 4. This agreement is also found in the remaining observations of the database. The data reduction method results in the elimination of the *inverted tunnel effect* in the background intensity obtained by the use of the standard data reduction tools. Early versions of these maps, with slight differences in the source detection and smoothing parameters in comparison to our method as well as a shorter accumulated exposure time, have been already successfully used for the detection of very faint extended X–ray emission in nearby dwarf galaxies by Kappes et al., in prep. However, it is important to note that this success is reduced to the inner region of the detector ($\sim$ CCD1), while we are interested in extending this analysis to the entire FOV.

Concerning the use of our exposure maps for specific XMM–Newton pointings, the maps must be re–scaled to the *effective* exposure time of the observation. In some cases, an additional re–scaling is necessary to adjust the overall shape of the exposure maps and the observed X–ray distribution before performing the correction for vignetting. This is because individual observations of the background – mainly in the low energy regime B1 and B2 – show real structure partly due to variations in the absorbing column density in the FOV and partly to the limitations in the source detection procedures which may fail to detect point sources with very faint emission. These effects occasionally result in an enhancement or, alternatively, in a decrease of the background intensity in the center region of the detector for the individual observations. Since this inner region ($r \leq 2$ arcmin) is used as default reference to normalize the re–scaling of the exposure maps, a new scaling (based in another detector region and generally within 10% of the default solution) might be needed in some cases to avoid an over- or underestimation of the total background count rate. At this point, the exposure maps are ready to be used.

The final results for our calculations for the *medium* filter case are shown in Figs. 5 to 7. There, the effect of bright CCD contamination (see next Sect.) has been almost completely eliminated and, therefore, the statistical significance of the final MOS1 exposure maps for the B1 and B2 energy bands is lower than for the remaining cases.

We also detected differences in the vignetting for alternative filters, as can be seen in Fig. 8. There, the radial profile of the exposure map for the MOS2 detector is compared for observations with the *thin*, *medium* and *thick* filters. Although in most of the cases the profiles are in agreement according to the error bars, some clearly deviate, like the profile for the *thick* filter in the B2 band or for *medium* in B4. Because of this differences we recommend the use of separate sets of exposure maps for each filter.

In the near future, electronic versions of the exposure maps presented in Figs. 5 to 7 (also for the cases of the *thin* and *thick* filters) will be available, together with dedicated software for the regridding of the maps from detector into the sky coordinates grid of a given XMM–Newton observation.

## 5. Radial Asymmetry of the XMM–Newton EPIC MOS1 Vignetting in the Soft Energy Regime

Until this point, we have supposed that the exposure maps of the EPIC detectors are radially symmetric. **The gaps of the so–called detector mask, the asymmetry at the rims of the FOV, and the presence of the Reflection Grating Arrays in the case of the MOS detectors, yield variations in the radial distribution of the exposure maps that argument in principle against the symmetry assumption. However, we added the data in our sample using always the same orientation (detector coordinates), which would unmask in the final result any important asymmetries originated in the mentioned effects. In our results, we find that the radial distribution of the exposure maps appears symmetric (dominated by the contribution of vignetting) for all bands** in the MOS2 camera, for bands B2 to B4 in pn and B3 to B4 in MOS1. All these exposure maps show their intensity maximum at the center of the FOV and the decrease with increasing radius is constant for all azimuth angles, like in the panels of Fig. 6. But this behavior is not observed in the exposure maps obtained for MOS1 in the B1 and B2 bands and the B1 band for pn. Since the asymmetries are larger in magnitude for the MOS1 case, we will focus on this detector in the following. The upper panel of Fig. 9 shows the results using all observations in the database for B1. Significant enhancements of the exposure map in two regions associated to CCDs 2 and 5 (numeration is adopted from the *XMM–Newton Users' Handbook*) are detected at a first visual inspection of the data. The size of the database that we have used – over 1.4 Ms observation time – and the number of contaminated observations, indicates a high statistical significance to the reality of these enhancements. In case of the B2 band, the intensity enhancement is confined to CCD5. In the following, we will focus on the investigation of these *bright CCDs* and, in particular, in the B1 regime.

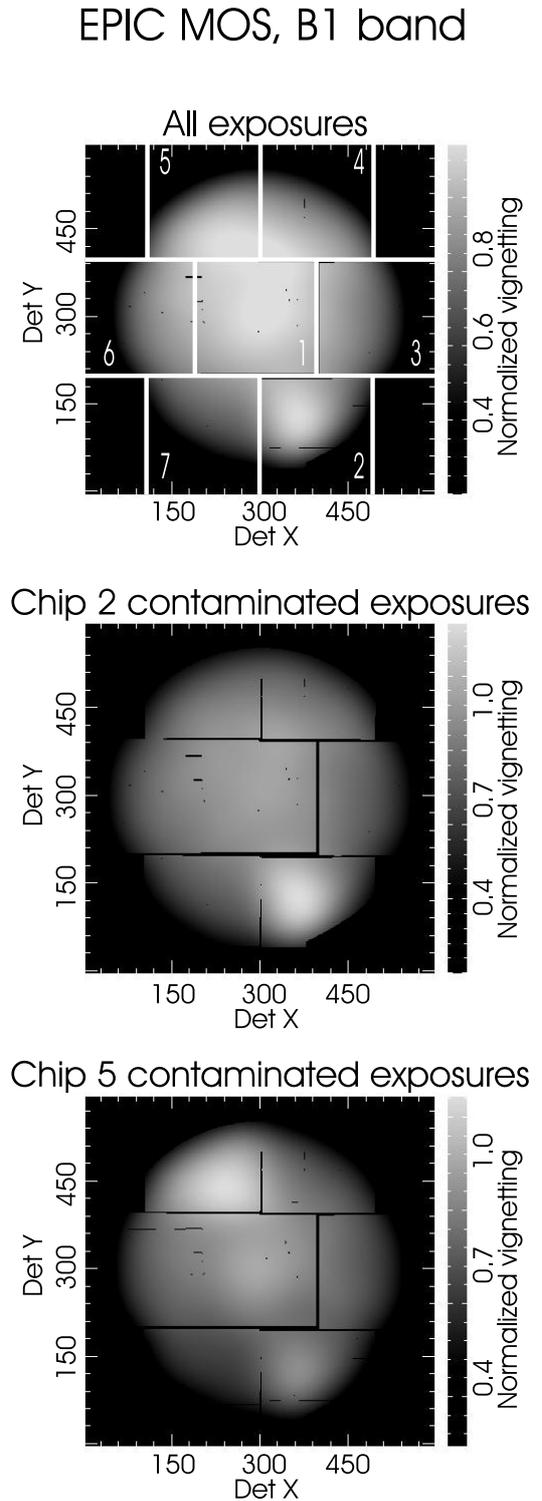

**Fig. 9.** EPIC MOS1 exposure maps calculated for the three separated data sets indicated in the top of each figure for the *medium* filter. The white grid and numbers shown in the **upper** panel indicate the positions of the seven CCDs in the MOS1 camera. The maps are normalized to a value of 1 in the center of the map. The count rate increase in the contaminated observations accounts up to a value of 1.5 in the affected region of the normalized exposure map. This corresponds up to a factor of 2 above the uncontaminated case (shown in the upper left panel of Fig. 5).

We performed an individual study of the observations in our database (all filters) and detected a number of observations which showed a clear enhancement, up to a factor of two, in the observed count rate in certain regions of the MOS1 detector, independently of the used filter. The count rate increase is confined to CCD2 in most cases. However, among these affected observations, some present an additional contamination of CCD5. In Tabs. 3 to 6, the superscripts [2] and [5] in the first column mark observations which present contamination, in CCD2 and 5 respectively. We selected as contaminated the observations with mean background intensities in CCD2 or CCD5 exceeding those of the uncontaminated CCDs by more than the level of accuracy of the data. More formally:

$$I_2 - \sigma_2 \geq I_{345} \qquad\qquad I_5 - \sigma_5 \geq I_{345} \qquad (1)$$

Here $I$ represents the mean background intensity, $\sigma$ the error in its determination and the sub–indexes represent the CCDs used to calculate the respective $I$ and $\sigma$ (see also Fig. 10 and Eq. 2).

It is important to note, that we also made use of more actual versions of the data analysis software (SAS 5.4.1) and calibration database (synchronization of March, 2004) to re–analyze the contaminated fields. The results obtained with the newest tools are only marginally different from those obtained with the older. The bright CCD contamination appears in the same manner in both cases. Furthermore, the use of the SAS task *eexpmap* to calculate the exposure maps in the contaminated observations yields results that do not reflect the intensity enhancement registered during the observations, **although the shape of the exposure maps calculated with the newest tools differs slightly from those obtained with the older versions**. To summarize, the standard tools do not perform any correction for the contamination reported here.

We also compared our set of affected observations with those analyzed by Read & Ponman (2003) to compute their EPIC background maps for the *medium* filter. From a total of 21 pointings in their sample, at least 3 contain CCD5 contamination (observation IDs 0021740101, 0022740301 and 008564021) and several more shown up with CCD2 contamination. The use of this contaminated observations produces an up to now undetected asymmetry in their MOS1 *event files*[4]

---

[4] http://www.sr.bham.ac.uk/xmm3/

and exposure maps for the soft energy regime (B1) equivalent to the one presented in Fig. 9.

After separating the observations in three sets, we repeated the calculation of the exposure maps for the cases: without bright CCDs, with bright CCD2 and with bright CCD5. The results for the contaminated sets are shown in the middle and lower panels of Fig. 9. With this first rough classification of the observations, the map obtained for the cases without bright CCDs (upper left panel in Fig. 5) is not completely satisfactory because the effect of CCD2 contamination is still present in the resulting exposure map. Furthermore, CCD2 contamination is also clearly present in the map calculated with the CCD5 contaminated observations (lower panel in Fig. 9). All this indicates that probably all observations are affected by the CCD2 contamination, as we might derive from the upper panels of Fig. 10. From those panels, we deduce that about 90% of the observations included in our sample show a higher background rate in CCD2 than in the remaining CCDs, although the excess is above the precision of the data only in about 50% of the cases (Eq. 1). In contrast, CCD5 contamination is confined to only a reduced fraction of the observations (15–20%) and its effect does not appear when calculating exposure maps with the remaining observations in the database: CCD2 contaminated and uncontaminated pointings in the middle and lower panels of Fig. 9 respectively. Another difference between both contaminations is given by the shape of the contaminated region in each case (see different shape for the contaminated regions in the middle and lower panels of Fig. 9). Additionally, the contamination of CCD5 extends into the B2 band, while the case of CCD2 is confined only to B1. All these considerations point to different origins for the contaminations of CCD2 and CCD5. In the following Sect., we present our first results in constraining these different origins. We will focus on energy band B1 because the bright CCD contamination in B2 is only present in 10% of the observations and never exceeds a factor 1.3 above the uncontaminated background intensity. However, the observations contaminated in the B2 band are excluded from the calculation of the final exposure map shown in the upper right panel of Fig. 5.

### 5.1. Quantification of the Contamination in MOS1 CCDs 2 and 5

In order to develop a reliable tool to correct for the intensity enhancement in CCDs 2 and 5, we searched for a quantification of this effect based on the measured difference between contaminated and uncontaminated observations. We define the *contamination rate* of CCD 2 $C_2$ as

$$C_2 = \frac{I_2}{I_{346}} \qquad (2)$$

Here, $I_2$ and $I_{346}$ represent the mean background intensities detected in CCD2 and in CCDs 3, 4 and 6 respectively (see also Eq. 1). We calculated the correlation coefficient of the contamination rate vs. different observation parameters like background count rate or effective exposure time. The upper panels of Fig. 10 show a graphical comparison of the quantities and Tab. 2 the results of the correlation analysis. An equivalent contamination rate is defined for CCD 5 as is shown in the lower panels of Fig. 10.

**Fig. 10. Upper panels:** Scatter plots of the CCD2 contamination as defined in Eq. 2 vs. the quantities indicated in the horizontal axes in logarithmic scale. The points with error bars represent contaminated observations (Eq. 1), where solid squares correspond to observations which also present CCD5 contamination. Empty squares represent uncontaminated observations. The curves correspond to the regression lines of the data. The statistical significance of the regression lines is shown in Tab. 2. The mean background intensities of the left panel are calculated using the data gained with CCDs 3, 4, 5 and 7. **Lower panels:** Equivalent to the upper panels for the CCD5 contamination. Points with error bars are CCD5 contaminated, solid squares show CCD2 contaminated observations and empty squares represent uncontaminated observations. No regression lines are shown because the statistical tests do not reject the zero correlation hypothesis in any of the three cases shown here (see Tab. 2 and Sect. 5.1). Visually, the CCD5 contaminated set is well separated from the remaining observations which are located close to the $C_5 = 1$ line as expected for pointings free of contamination.

CCD2 contaminated observations (points with error bars in the upper panels of Fig. 10) are concentrated towards the high background count rate part of the diagram. There is also a clear tendency to present higher contamination rates for higher count rates. This is confirmed by the statistical test for the significance of the correlation coefficient of the sample, which rejects the zero correlation hypothesis between contamination rate (Eq. 2) and the background count rates (see Tab. 2). With respect to the effective exposure time, after proton flare filtering, a negative correlation can be rejected at a 95% significance level but not at a 99% level. For this case, a larger sample of contaminated observations is necessary to derive definitive statistical conclusions. The correlation of CCD2 contamination rate with total counts is also accepted by the test, but the reliability of the correlation is significantly lower with respect to the case of the correlation with the background count rate. Furthermore, there are also uncontaminated observations in the high effective exposure time and high total counts parts of the diagrams (different to the count rate case). This indicates that probably the correlations found between the contamination rate and these two parameters (effective exposure time and total counts) are not real. As already mentioned, more data are necessary to derive definitive conclusions in this sense.

For the CCD5 contaminated data set, we detect a systematic overestimation in the CCD5 intensities of about a factor of two (points with error bars in the lower panels of Fig. 10). In contrast to the case of CCD2, there is no evidence for grouping of the contaminated observations towards any count rate, exposure time or total counts regime for CCD5 contamination. The correlation coefficients shown in the right column of Tab. 2 do not reject the un–correlation in any of the three cases with contamination rate.

We also tested whether the bright CCD contamination is triggered by extreme values in the X–ray background hardness ratios of the observations or is correlated with the exposure date (see Fig. 11 and Tabs. 3 to 6) and no relation could be found.

**Fig. 11.** Hardness ratios of the X–ray backgrounds as detected with the EPIC MOS1 camera for the observations in Tabs. 3 to 6 (all filters). Empty squares correspond to observations free of CCD2 or CCD5 contamination, solid squares to CCD2 contaminated observations and the points with the error bars to CCD5 contaminations. However, in order to avoid the influence of this contamination in the hardness ratios presented here, we calculate the values based only on the data provided by uncontaminated CCDs. The results shown in this figure rule out a correlation between the hardness ratio of the X–ray background towards the observed field and the presence of bright CCD contamination, either for the CCD5 or for the CCD2 case. This can be deduced from the fact that the contaminated observations are spread in the panel with a similar distribution to that of the uncontaminated observations.

**Table 2.** Correlation coefficients of the CCD contamination rates as defined in Sect. 5 vs. the observation parameters listed in the left column of the table. The last row shows the critical value of a the statistical test for the significance of the correlation coefficient of the sample at a 95% confidence level adopted from Crow et al. (1960). We shown the 99% confidence level value in parenthesis. Note that the zero correlation is only rejected with the values of the CCD2 column.

|                | $C_2$         | $C_5$         |
| -------------- | ------------- | ------------- |
| Count rate     | 0.806         | −0.127        |
| Effective time | −0.297        | −0.338        |
| Total counts   | 0.541         | −0.246        |
| $|r|_{crit}$   | 0.235 (0.305) | 0.413 (0.526) |

From our analysis of the bright CCD contaminations, we conclude that they probably have their origin in *software* problems or in the electronics of the CCD. The contamination is always present and grows with mean background count rate in the case of CCD2. Therefore, the use of diffuse X–ray data from this CCD2 should be avoided until a consistent method to mitigate the effect of the contamination is developed. For CCD5 case, since the frequency in the presence of this contamination is below 20% and the contamination rate is always high ($C_5 > 1.5$ using an equivalent definition to that of Eq. 2), we recommend to check whether the intensity increment is visible in the raw data before performing further analysis of the diffuse X–ray data gained with this CCD.

## 6. Conclusions

Our study of the in–flight performance of XMM–Newton has produced new results concerning the treatment of *proton flares* and *vignetting*. These results imply the necessity of the inclusion of the soft energy bands in the search for proton flares, different to the conclusions of previous works (Lumb et al. 2002; Marty et al. 2003; Read & Ponman 2003). For this data reduction step, we developed an iterative automatic procedure that can be "pipelined" to the XMM–Newton raw data.

Since the standard tools for the vignetting correction in the XMM–Newton EPIC detectors are not sufficient to allow a sensitive analysis of faint signals in the soft energy regime ($E \leq 2.0$ keV), as we can see in Fig. 3, we have developed a new task to calculate the *real* exposure map of XMM–Newton. It is based on the in–flight performance of the observatory and uses longest accumulated exposure time today. Our results point to differences between the vignetting of observations performed with different filters. The process to calculate exposure maps can be repeated automatically for user defined energy bands and observations. An extension to modes alternative to the *FullFrame* mode presented here can be performed by exploiting the XMM–Newton Science Archive.

We detect an artificial increase in the measured count rates (softest energy, $0.2$ keV $\leq E \leq 0.5$ keV) in the EPIC MOS1 CCDs 2 and 5. This contamination appears to be present in practically all observations and it seems to be associated to high background rates in the case of CCD2. The set of affected observations accumulated for CCD5 contamination until today is not sufficient to draw definitive conclusions on the origin of the contamination. The count rate in the affected CCDs is increased up to a factor of two. Such a significant difference must be taken into account when investigating diffuse emission or the hardness ratios of faint sources detected by one of these two CCDs. Also, the EPIC pn exposure map presents asymmetries in the lower energy regime, which are and obstacle for use of the outer parts of the detector to investigate faint diffuse emission. However, the asymmetries in the pn case are lower in magnitude that those for the MOS1 detector.

To summarize, the actual status of the data reduction for XMM–Newton still presents fundamental problems which prevent a reliable study of faint soft X–ray sources (e.g., *Bright CCD* contamination). These difficulties have to be further investigated and solved before being able of using the full capabilities of the satellite. Interesting topics that could be in principle investigated with XMM–Newton, like for example:

– emissivity distribution of faint X–ray halos of external galaxies,
– the X–ray emission of the WHIM,
– the degree of clumping of the ISM in its X–ray emitting and absorbing phases.

can be only investigated at present, if the X–ray emission is concentrated in the inner part of the FOV ($r < 5$ arcmin),

where the uncertainties in the data reduction are reduced. This excludes the study of the three topics mentioned above which involve, in general, structures covering significantly larger angular scales.

*Acknowledgements.* JP likes to thank the Deutsches Zentrum für Luft- und Raumfahrt and the Graduiertenkolleg 787 of the universities of Bonn and Bochum for the financial support.

**Table 3.** Summary of the observations extracted from the XMM–Newton Science Archive for our investigation of the XMM–data reduction. This table shows observations performed with the *thin* filter. The fourth column shows the total column densities derived from the newest all–sky stray–radiation corrected 21–cm line surveys (Kalberla et al., submitted). The $t_{total}$ columns show the total integration time of the observation for the MOS and pn instruments. The $t_{effective}$ columns show the effective time after proton flare filtering (see Sect. 3.2) for each observation in MOS1, MOS2 and pn as indicated. The accumulated integration time of the *thin* filter sample is $t_{total} \simeq 1.6$ Ms for MOS and $t_{total} \simeq 1.4$ Ms for pn. The accumulated effective time amounts to $t_{eff} \simeq 1.25$ Ms for MOS and $t_{eff} \simeq 1.1$ Ms for pn.

| Observation ID | Galactic coordinates | | $N_{H\,I}$ [$10^{20}$ cm$^{-2}$] | $t_{total}$ [ks] | | $t_{effective}$ [ks] | | |
|---|---|---|---|---|---|---|---|---|
| | l [°] | b [°] | | MOS | pn | MOS1 | MOS2 | pn |
| 0001930101 | 112.9272 | -51.6857 | 4.2 | 23.8 | 21.4 | 16.2 | 15.9 | 16.0 |
| 0025740401 | 68.3137 | 48.1185 | 0.9 | 16.8 | 21.5 | 9.2 | 10.0 | 7.4 |
| 0026340101 | 140.7876 | 38.6539 | 2.5 | 26.0 | 22.0 | 25.5 | 25.8 | 21.7 |
| 0033540901 | 60.4069 | 42.9379 | 1.0 | 17.0 | 14.5 | 13.5 | 0.4 | 9.7 |
| 0036540101[2] | 185.5553 | -41.3712 | 6.1 | 22.3 | 20.0 | 18.6 | 19.7 | 18.0 |
| 0038540301 | 36.8118 | -35.0917 | 3.2 | 16.6 | 14.8 | 16.6 | 16.6 | 14.8 |
| 0051760201[2] | 21.5406 | 72.3636 | 2.3 | 20.8 | 18.3 | 16.9 | 16.8 | 18.3 |
| 0056021601 | 46.8803 | -56.6305 | 3.9 | 24.4 | 20.0 | 18.6 | 18.3 | 16.2 |
| 0081340401[2] | 358.7007 | -40.8013 | 3.2 | 21.1 | 18.5 | 13.8 | 12.7 | 11.5 |
| 0081341001[2] | 322.5620 | -28.7663 | 6.1 | 22.6 | 19.9 | 16.3 | 16.7 | 14.4 |
| 0081341101 | 57.8763 | 55.9408 | 1.5 | 19.4 | 17.1 | 19.2 | 19.1 | 14.8 |
| 0083240201[2] | 224.7495 | 33.6648 | 2.7 | 20.1 | 15.7 | 17.3 | 18.6 | 14.4 |
| 0086770101 | 280.7828 | -32.5647 | 19.0 | 46.4 | 44.0 | 41.7 | 39.1 | 44.0 |
| 0092360801 | 302.6432 | -20.3869 | 8.6 | 16.1 | 11.4 | 16.1 | 15.6 | 10.8 |
| 0094530401[2,5] | 197.2746 | 26.6316 | 2.9 | 24.3 | 20.0 | 19.6 | 19.7 | 9.9 |
| 0094800201[2] | 133.9344 | 49.4457 | 1.0 | 53.0 | 50.2 | 24.8 | 24.9 | 24.1 |
| 0098810101[2] | 272.2947 | -58.1071 | 3.3 | 23.5 | 21.1 | 21.3 | 21.5 | 20.1 |
| 0099030101 | 230.9615 | 66.4559 | 1.3 | 22.3 | 19.9 | 14.5 | 14.4 | 10.5 |
| 0106660101 | 39.3340 | -52.9305 | 1.7 | 57.6 | 55.2 | 54.0 | 55.7 | 53.9 |
| 0106660601[2,5] | 39.3340 | -52.9305 | 1.7 | 109.4 | 106.8 | 92.6 | 92.2 | 78.0 |
| 0106660201[2] | 39.3340 | -52.9305 | 1.7 | 52.5 | 50.1 | 45.8 | 42.5 | 37.0 |
| 0109520301 | 172.4940 | -58.3013 | 2.2 | 22.0 | 18.0 | 21.1 | 20.8 | 17.8 |
| 0109520501 | 171.5545 | -58.7493 | 2.0 | 24.2 | 20.0 | 22.9 | 23.0 | 19.9 |
| 0109520601 | 170.8890 | -58.6112 | 2.0 | 23.0 | 19.0 | 22.0 | 22.6 | 19.0 |
| 0110980101 | 240.9164 | 64.7794 | 1.9 | 54.5 | 50.7 | 47.5 | 46.8 | 41.8 |
| 0110980401 | 281.4384 | -40.7275 | 6.2 | 43.8 | 40.0 | 32.7 | 31.4 | 26.7 |
| 0110990201[2] | 289.2767 | 63.7165 | 1.9 | 28.9 | 24.7 | 25.3 | 25.6 | 22.0 |
| 0111110101 | 173.6176 | -58.2030 | 2.1 | 25.2 | 20.9 | 16.1 | 16.4 | 11.3 |
| 0111110401 | 172.2250 | -58.8840 | 2.5 | 28.5 | 22.5 | 27.4 | 26.6 | 21.0 |
| 0111110501 | 171.7491 | -59.1079 | 2.2 | 24.3 | 20.1 | 18.4 | 21.1 | 13.2 |
| 0112230901[2] | 6.6178 | 30.7039 | 11.1 | 27.2 | 23.9 | 10.2 | 9.5 | 22.0 |
| 0112231001 | 6.5454 | 30.2856 | 11.5 | 28.4 | 24.5 | 24.2 | 24.1 | 21.0 |
| 0112250301 | 358.7947 | 64.7724 | 1.9 | 25.7 | 22.0 | 24.7 | 24.0 | 21.2 |
| 0112370101[2,5] | 169.8191 | -59.7521 | 2.0 | 57.6 | 55.3 | 38.7 | 36.8 | 32.2 |
| 0112370301[2,5] | 170.4148 | -59.4912 | 1.9 | 63.1 | 60.7 | 39.2 | 39.3 | 26.0 |
| 0112371001 | 169.8191 | -59.7521 | 2.0 | 59.7 | 57.5 | 38.6 | 38.7 | 33.6 |
| 0112480101[2,5] | 355.5732 | 14.6076 | 11.5 | 20.1 | 15.8 | 18.0 | 19.5 | 13.0 |
| 0112480201 | 355.3334 | 14.4177 | 11.8 | 18.6 | 14.3 | 18.0 | 18.4 | 14.0 |
| 0112480301[2,5] | 355.0940 | 14.2275 | 11.8 | 18.0 | 14.0 | 17.3 | 14.6 | 10.0 |
| 0112650401 | 132.0265 | -69.0433 | 6.2 | 23.8 | 20.0 | 19.2 | 23.4 | 15.2 |
| 0112680401[2] | 172.7554 | -57.7180 | 2.2 | 24.3 | 22.0 | 21.7 | 21.9 | 21.7 |
| 0112810201[2] | 277.4331 | 11.7246 | 9.2 | 16.5 | 14.5 | 6.2 | 5.3 | 3.2 |
| 0124900101[2] | 126.5655 | 52.7495 | 2.5 | 54.6 | 55.7 | 54.6 | 54.7 | 55.7 |
| 0127920401[2] | 96.4360 | 60.0775 | 0.9 | 30.2 | 26.7 | 13.8 | 15.3 | 15.5 |
| 0127921001 | 96.4360 | 60.0775 | 0.9 | 55.9 | 51.5 | 52.3 | 52.5 | 48.5 |
| 0127921201 | 96.4360 | 60.0775 | 0.9 | 18.5 | 14.7 | 17.1 | 17.8 | 14.7 |
| 0128530301[2] | 237.5013 | 16.3102 | 5.7 | 37.5 | 8.8 | 31.1 | 32.0 | 7.4 |
| 0136040101 | 35.0105 | 49.1550 | 4.1 | 26.0 | 23.7 | 16.4 | 17.4 | 15.2 |
| 0137750101[2] | 303.8713 | 3.3423 | 57.1 | 18.9 | 16.4 | 16.5 | 16.3 | 14.4 |

[2,5]: intensity enhancement in EPIC MOS 1 CCD 2 and 5 respectively (see Sect. 5)

**Table 4.** Like Tab. 3 for the case of the observations with the *medium* filter. The accumulated integration time of this sample is $t_{total} \simeq 1.4$ Ms for MOS and $t_{total} \simeq 1.3$ Ms for pn. The accumulated effective time amounts to $t_{eff} \simeq 1.1$ Ms for MOS and $t_{eff} \simeq 1.0$ Ms for pn.

| Observation ID | Galactic coordinates | | $N_{HI}$ | $t_{total}$ [ks] | | $t_{effective}$ [ks] | | |
|---|---|---|---|---|---|---|---|---|
| | l [°] | b [°] | [$10^{20}$ cm$^{-2}$] | MOS | pn | MOS1 | MOS2 | pn |
| 0000110101[2,5] | 149.2383 | 4.1332 | 36.0 | 32.3 | 29.7 | 21.5 | 21.7 | 19.6 |
| 0007420701 | 311.1053 | -0.4177 | 210.4 | 12.1 | 9.7 | 9.9 | 11.9 | 9.7 |
| 0007420801 | 310.8963 | -0.0011 | 207.2 | 13.0 | 12.4 | 12.0 | 12.8 | 12.4 |
| 0007421001 | 310.8963 | 0.8323 | 182.1 | 12.0 | 9.6 | 9.5 | 7.4 | 4.5 |
| 0007421901[2] | 311.3133 | 0.8323 | 177.0 | 10.6 | 8.0 | 9.6 | 9.1 | 8.0 |
| 0007422001 | 311.5213 | 1.2489 | 145.6 | 10.0 | 7.6 | 10.0 | 9.2 | 7.6 |
| 0007422101 | 311.3133 | 1.6656 | 118.8 | 11.8 | 9.4 | 11.8 | 11.4 | 9.4 |
| 0007422201 | 311.5213 | 2.0823 | 83.4 | 11.8 | 9.4 | 11.8 | 11.8 | 9.4 |
| 0007422301 | 311.3133 | 2.4989 | 91.2 | 15.0 | 12.6 | 13.7 | 14.8 | 12.6 |
| 0021740101[2,5] | 188.4691 | 48.6591 | 0.9 | 34.3 | 30.0 | 28.8 | 28.7 | 21.6 |
| 0022140101[2] | 336.4291 | -0.2192 | 196.6 | 15.9 | 11.2 | 15.9 | 13.2 | 11.2 |
| 0022740101[2,5] | 149.3420 | 53.1450 | 0.6 | 83.4 | 80.8 | 54.3 | 51.8 | 54.8 |
| 0022740201[2,5] | 149.3420 | 53.1450 | 0.6 | 63.9 | 61.6 | 45.5 | 45.0 | 42.1 |
| 0022740301[2,5] | 149.3420 | 53.1450 | 0.6 | 38.0 | 36.9 | 34.4 | 32.0 | 27.4 |
| 0026340201[2,5] | 246.2280 | 39.8908 | 3.8 | 19.1 | 15.0 | 9.4 | 9.3 | 14.1 |
| 0026340301 | 140.2725 | 43.6037 | 3.1 | 24.0 | 20.0 | 20.0 | 20.1 | 12.9 |
| 0032140201 | 82.7260 | -45.5733 | 3.8 | 12.5 | 10.0 | 6.9 | 7.0 | 5.9 |
| 0046940401[2,5] | 216.4227 | 45.5089 | 2.7 | 15.0 | 10.6 | 14.8 | 14.1 | 10.6 |
| 0050940101 | 1.1080 | -3.8737 | 29.3 | 24.0 | 20.0 | 9.4 | 8.8 | 3.1 |
| 0050940301[2] | 0.6202 | -8.0014 | 13.7 | 13.5 | 9.1 | 9.8 | 9.3 | 7.5 |
| 0051610101[2] | 345.0384 | -27.7477 | 3.9 | 22.0 | 18.0 | 16.7 | 16.9 | 12.7 |
| 0052140201[2,5] | 174.7759 | 68.4919 | 2.1 | 40.6 | 36.3 | 29.3 | 28.9 | 22.1 |
| 0058940101 | 137.5811 | 35.5431 | 1.7 | 27.8 | 23.8 | 26.3 | 26.9 | 23.6 |
| 0058940301[2] | 319.8064 | 26.4422 | 5.5 | 19.3 | 15.3 | 18.8 | 19.1 | 15.3 |
| 0067340201[2] | 350.0626 | -10.0013 | 9.5 | 14.6 | 9.9 | 12.0 | 12.3 | 8.4 |
| 0070340301[2] | 185.2610 | 65.4974 | 2.9 | 31.4 | 32.8 | 24.6 | 25.1 | 21.0 |
| 0070341201 | 97.0904 | 42.6041 | 1.6 | 22.2 | 15.2 | 21.5 | 22.0 | 14.6 |
| 0079570201[2,5] | 234.5553 | -10.1417 | 19.1 | 47.6 | 43.5 | 32.4 | 32.3 | 30.5 |
| 0082140301 | 241.3907 | 64.2189 | 2.0 | 32.9 | 28.9 | 31.4 | 30.1 | 28.9 |
| 0083950101[2] | 161.4756 | -13.6351 | 13.7 | 27.3 | 23.3 | 21.7 | 21.1 | 16.6 |
| 0085640201[2,5] | 152.5383 | 42.8960 | 3.1 | 34.4 | 30.0 | 33.7 | 33.8 | 24.5 |
| 0092800201 | 165.8166 | 36.2395 | 4.2 | 93.8 | 100.0 | 68.6 | 69.8 | 100.0 |
| 0092970201 | 279.1716 | -64.5343 | 1.8 | 13.6 | 9.0 | 13.6 | 12.6 | 8.2 |
| 0093640701[2] | 348.2068 | -65.2380 | 1.4 | 20.0 | 15.4 | 18.5 | 17.9 | 15.4 |
| 0093640901 | 138.2363 | 10.5812 | 32.3 | 9.8 | 5.9 | 9.8 | 7.7 | 2.1 |
| 0093670501 | 347.3321 | -0.4998 | 148.1 | 14.1 | 9.2 | 13.3 | 13.8 | 9.2 |
| 0100640101[2,5] | 122.8397 | 22.4717 | 6.2 | 43.3 | 40.9 | 26.8 | 26.9 | 22.7 |
| 0102640201 | 133.8045 | -30.9986 | 13.4 | 17.1 | 13.3 | 15.6 | 15.4 | 9.5 |
| 0104460301[2] | 20.0633 | -0.0024 | 162.0 | 12.0 | 9.6 | 4.5 | 12.0 | 6.9 |
| 0109110101[2] | 297.6201 | 0.3362 | 132.0 | 76.0 | 72.0 | 69.0 | 69.8 | 63.8 |
| 0111120201[2] | 353.0714 | 16.5547 | 12.3 | 32.8 | 30.4 | 26.4 | 24.2 | 26.0 |
| 0112190101 | 162.8518 | -34.8370 | 8.8 | 13.4 | 10.0 | 12.3 | 13.4 | 6.5 |
| 0112190201 | 155.2194 | 75.3140 | 2.6 | 14.0 | 10.0 | 14.0 | 13.3 | 10.0 |
| 0112190401[2] | 6.6839 | 43.0968 | 4.4 | 14.4 | 10.0 | 13.0 | 13.8 | 10.0 |
| 0112220101[2] | 309.1624 | 14.9714 | 8.2 | 40.0 | 37.5 | 37.1 | 39.4 | 36.6 |
| 0112500101 | 111.6403 | 31.9333 | 4.1 | 25.4 | 22.0 | 24.5 | 24.6 | 22.0 |
| 0112970201[2] | 0.9315 | 0.0820 | 122.2 | 17.4 | 13.5 | 14.2 | 17.4 | 10.2 |
| 0112970701 | 359.6176 | -0.0468 | 124.1 | 23.9 | 20.0 | 23.3 | 23.9 | 20.0 |
| 0129320801 | 270.2126 | -51.6390 | 1.1 | 10.0 | 6.2 | 9.7 | 10.0 | 6.2 |
| 0135740901[2] | 21.0629 | 0.3448 | 188.2 | 11.9 | 9.2 | 8.8 | 9.0 | 6.9 |

[2] like in Tab. 3

**Table 5.** Like Tab. 3 for the case of the observations with the *thick* filter. The accumulated integration time of this sample is $t_{\rm total} \simeq 0.77$ Ms for MOS and $t_{\rm total} \simeq 0.73$ Ms for pn. The accumulated effective time amounts to $t_{\rm eff} \simeq 0.65$ Ms for MOS and $t_{\rm eff} \simeq 0.63$ Ms for pn.

| Observation ID | Galactic coordinates | | $N_{\rm H\,I}$ | $t_{\rm total}$ [ks] | | $t_{\rm effective}$ [ks] | | |
|---|---|---|---|---|---|---|---|---|
| | l [°] | b [°] | [$10^{20}$ cm$^{-2}$] | MOS | pn | MOS1 | MOS2 | pn |
| 0006010301 | 192.2911 | 23.4055 | 4.1 | 34.6 | 32.0 | 33.0 | 33.2 | 30.6 |
| 0006010401 | 304.8355 | -39.7799 | 4.0 | 35.3 | 33.0 | 33.3 | 34.9 | 32.5 |
| 0021750501 | 321.5983 | -15.2598 | 8.3 | 28.1 | 24.7 | 23.0 | 21.5 | 19.6 |
| 0021750701 | 321.5983 | -15.2598 | 8.3 | 28.3 | 25.6 | 23.7 | 25.2 | 25.6 |
| 0024140101[2,5] | 179.1452 | -23.8159 | 13.6 | 61.1 | 58.4 | 50.9 | 61.1 | 58.4 |
| 0044740201[2] | 270.5877 | 60.7549 | 2.3 | 48.3 | 45.8 | 41.6 | 45.8 | 45.8 |
| 0100440101 | 59.1965 | -49.6051 | 4.7 | 45.8 | 43.3 | 40.6 | 41.8 | 41.4 |
| 0109120101 | 117.9887 | 1.2514 | 60.8 | 36.8 | 33.0 | 34.4 | 33.6 | 33.0 |
| 0109260201 | 277.3024 | -36.1355 | 13.5 | 32.9 | 28.5 | 32.7 | 31.4 | 28.5 |
| 0109280101[2] | 292.3783 | -4.8299 | 36.3 | 23.9 | 19.9 | 20.6 | 20.4 | 16.2 |
| 0112551501[2] | 147.9399 | -54.1548 | 5.0 | 21.2 | 17.8 | 14.6 | 14.7 | 17.0 |
| 0112880401[2,5] | 197.6973 | -23.7610 | 6.2 | 19.0 | 17.0 | 16.7 | 19.0 | 17.0 |
| 0113891001[2] | 274.0034 | -15.8817 | 12.6 | 20.0 | 20.0 | 18.2 | 17.7 | 18.0 |
| 0113891101[2,5] | 274.0034 | -15.8817 | 12.6 | 17.0 | 17.0 | 13.4 | 9.8 | 10.6 |
| 0122520201 | 293.5047 | -37.5515 | 8.3 | 36.4 | 36.4 | 36.4 | 36.4 | 36.4 |
| 0123700401[2] | 149.3420 | 53.1450 | 0.6 | 15.0 | 16.5 | 12.1 | 14.3 | 8.3 |
| 0126511201 | 273.9041 | -15.8264 | 11.8 | 27.5 | 29.2 | 19.6 | 20.4 | 16.4 |
| 0134531201 | 273.9041 | -15.8264 | 11.8 | 21.4 | 19.0 | 19.2 | 21.4 | 16.6 |
| 0134531501[2,5] | 273.9041 | -15.8264 | 11.8 | 21.2 | 18.6 | 20.4 | 18.2 | 16.5 |
| 0140160101[2] | 143.7135 | -15.5579 | 6.3 | 39.5 | 41.8 | 21.0 | 20.6 | 16.6 |
| 0142630301 | 347.1868 | 21.5108 | 10.5 | 22.0 | 20.3 | 17.4 | 17.8 | 14.9 |
| 0143370101[2] | 217.3147 | -28.9072 | 4.5 | 47.0 | 45.4 | 38.6 | 38.1 | 45.4 |
| 0148680101[2] | 43.5776 | 85.4061 | 1.3 | 61.4 | 63.8 | 47.5 | 45.2 | 63.8 |
| 0205650101[2] | 107.1315 | -0.8986 | 58.8 | 30.4 | 27.6 | 7.3 | 7.9 | 4.0 |

[2] like in Tab. 3

**Table 6.** Summary of of our sample of pointings performed with XMM–Newton (see also Tab. 3). The directions of the pointings have been specifically selected to study the contribution of the SXRB as seen by XMM–Newton.

| Observation ID | Galactic coordinates | | $N_{\rm H\,I}$ | $t_{\rm total}$ [ks] | | $t_{\rm effective}$ [ks] | | |
|---|---|---|---|---|---|---|---|---|
| | l [°] | b [°] | [$10^{20}$ cm$^{-2}$] | MOS | pn | MOS1 | MOS2 | pn |
| 0110660301 | 125.0554 | 30.5087 | 8.2 | 8.9 | 5.2 | 5.4 | 5.7 | 2.9 |
| 0110660401 | 165.1789 | 66.5823 | 3.3 | 13.2 | 9.3 | 11.1 | 10.9 | 6.0 |
| 0110660601[2] | 95.0340 | 38.9844 | 2.6 | 16.1 | 12.4 | 12.0 | 13.6 | 12.2 |
| 0110660801[2] | 91.3173 | 37.0343 | 2.3 | 12.2 | 9.2 | 10.3 | 10.4 | 9.2 |
| 0110661601[2] | 125.0554 | 30.5087 | 8.2 | 10.0 | 6.0 | 10.0 | 10.0 | 6.0 |
| 0110661701[2] | 153.2919 | 39.3394 | 6.4 | 11.2 | 11.6 | 9.7 | 11.2 | 11.6 |
| 0110662401[2] | 87.9566 | 36.3392 | 2.5 | 8.7 | 5.6 | 8.7 | 8.7 | 5.6 |
| 0110662601[2] | 90.8139 | 37.9770 | 1.7 | 11.4 | 8.0 | 10.1 | 9.8 | 8.0 |
| 0110662701[2] | 88.9394 | 37.1270 | 1.9 | 9.7 | 6.3 | 7.9 | 8.0 | 4.9 |

[2] like in Tab. 3